\documentclass[aps,pra,epsfigure,twocolumn,showpacs,superscriptaddress]{revtex4}
\usepackage{color}
\usepackage{dcolumn}    
\usepackage{bm} 
\usepackage{graphicx}
\usepackage{amsmath}    
\usepackage{latexsym}
\usepackage{amsfonts}   
\usepackage{amssymb}
\usepackage{array}      
\usepackage{epsfig}
\usepackage{txfonts}
\usepackage{ulem}

\newcommand{\ket}[1]{\left\vert#1\right\rangle}
\newcommand{\bra}[1]{\left\langle#1\right\vert}

\begin{document}
\title{Correlation approach to work extraction from  finite quantum systems}
\author{Gian Luca Giorgi}
\affiliation{INRIM, Strada delle Cacce 91, I-10135 Torino, Italy}
\email{g.giorgi@inrim.it}
\author{Steve Campbell}
\affiliation{Centre for Theoretical Atomic, Molecular and Optical Physics, School of Mathematics and Physics, Queen's University, Belfast BT7 1NN, United Kingdom}
\email{steven.campbell@qub.ac.uk}

\begin{abstract}

Reversible work extraction from identical quantum systems via collective operations was shown to be possible even without producing entanglement among the sub-parts. Here, we show that implementing such global operations necessarily imply the creation of quantum correlations, as measured by quantum discord. 
We also reanalyze the conditions under which global transformations outperform local gates as far as maximal work extraction is considered by deriving a necessary and sufficient condition that is  based on classical correlations.

\end{abstract}
\date{\today}
\maketitle

\section{Introduction}
Determining the maximum amount of work that can be extracted from a system by means of a cyclical transformation is one of the central problems in thermodynamics. In the context of finite quantum systems, the solution to this problem was found in Ref. \cite{Allahverdyan}. It was shown that thermally isolated finite systems are less efficient than macroscopic ones. This can be traced back to the fact that, in the absence of dissipation, not only does the entropy need to be conserved during the cycle, but also all the eigenvalues of the density operator. The problem of work extraction from a small quantum system was also discussed in Refs. \cite{horodecki,popescu}.

More recently, the way the presence of entanglement affects maximal work extraction was investigated in Refs. \cite{alicki,hovna}.  Alicki and Fannes proved that given $n$ identical copies of the system, nonlocal unitary operations are capable of increasing the amount of work extracted with respect to local operations \cite{alicki}. As clarified by Hovhannisyan \textit{et al.},  nonlocal operations do  not necessarily imply entanglement generation  \cite{hovna}. In fact, there exist regions in the system and Hamiltonian parameters where, even if the system remains separable at any time, maximal work extraction is reached. That being said, generating entanglement is unavoidable when the amount of work extracted is very high. Following a different approach, it had also been shown that entanglement is not a fundamental ingredient for work extraction from a heat bath using bipartite states, the essential resource being the so-called work deficit, a legitimate candidate to quantify quantum correlation \cite{Oppenheim}.

Entanglement is a distinctive feature of quantum mechanics, as its role has been shown to be fundamental in different quantum computation and communication contexts. However, there are tasks  where quantum advantages can be reached even using separable states. Prominent examples are given by the so-called deterministic quantum computation with one qubit (DQC1) protocol \cite{knill,datta,lanyon}, quantum state discrimination \cite{qsd}, and remote state preparation \cite{rsp}, where the description of quantumness beyond entanglement employed is the quantum discord (QD)~\cite{Ollivier,Henderson}. However, in these examples the exact ``role" played by the QD is unclear. Indeed, of the resource nature of QD is still an open question, and there have been recent advances showing that the presence of QD does not necessarily imply any better than classical advantage~\cite{terno,pra}. That said, the use of QD to study quantum phenomena and to understand the limitations of a quantum system to perform certain useful protocols has been very fruitful~\cite{gu,distribution}. In this paper we examine a protocol for extracting work from an array of quantum batteries. 
We show that, using collective operations, it is not possible to extract the maximum amount of work without producing quantum correlations during the dynamics, that is, unlike entanglement, the generation of discord is necessary in general for maximal work extraction. Our investigation is somewhat related to the results of Ref. \cite{Huber14}, where, considering the opposite point of view, the authors studied the limitations imposed by thermodynamics for creating correlations.

The remainder of the paper is organized as follows, we will start with a brief review the definitions of quantum discord and its generalization to multipartite systems in Sec. \ref{sec:discord}. In Sec. \ref{sec:work} we will introduce the model and discuss the different strategies that can be implemented in order to maximize the work extracted. In Sec. \ref{owe} we present our main results: we discuss two pedagogical cases of work extraction from two qubits and two qutrits, after which we generalize the results to $n$-qudits and show that maximal work extraction cannot be implemented without creating discord. Finally, conclusions are given in Sec. \ref{sec:conclusions}.

\section{Quantum discord in bipartite and multipartite systems}
\label{sec:discord}
In this section we define the three variations of quantum discord considered throughout the paper, the ``standard" quantum discord (QD), the global discord (GD), and the genuine quantum correlations (GC).

\subsection{Quantum discord}
We begin by recapitulating the definition of quantum discord (QD)~\cite{Ollivier}. Put simply, it is a measure of the quantum nature of a given bipartite state quantified by the difference between two quantum versions of the information content that are classically equivalent. Given a bipartite system $\rho_{AB}$ with $\rho_{A}$ ($\rho_{B}$) the reduced state of system $A$ ($B$), the mutual information is then
\begin{equation}
I(\rho_{AB})=S(\rho_{A})-S(\rho_{A}|\rho_{B}),
\label{MI}
\end{equation}
with $S(\rho_{A})=-\mathrm{Tr}[\rho_{A}\log_2\rho_{A}]$ the von Neumann entropy and $S(\rho_{A}|\rho_{B})=S(\rho_{AB})-S(\rho_B)$ is the conditional entropy. A classically equivalent expression can be formulated using a measurement-based approach. Allowing for a local projective measurement, described by the set of projectors $\{\hat{\Pi}_B^j\}$ on B, we arrive at the conditional post-measurement density operator $\rho_{AB|j}=(\openone_A\otimes\hat{\Pi}_B^j)\rho_{AB}(\openone_A\otimes\hat{\Pi}_B^j)/p_j$, where $p_j=\mathrm{Tr}[(\openone_A\otimes\hat{\Pi}_B^j)\rho_{AB}]$ is the probability associated with the measurement outcome $j$. The measurement-based conditional entropy $S(\rho_{AB}|\hat{\Pi}_B^j)=\sum_j p_j S(\rho_{A|j})$ with $\rho_{A|j}=\mathrm{Tr}[\hat{\Pi}^j_B\rho_{AB}]/p_j$ leads us to the one-way classical information~\cite{Henderson} 
\begin{equation}
J(\rho_{AB})=S(\rho_A)-S(\rho_{AB}|\hat{\Pi}_B^j).
\label{CI}
\end{equation}
The difference between Eqs~(\ref{MI}) and (\ref{CI}) is then minimized over the whole set of POVM's performed on B. Thus the QD is defined
\begin{equation}
\label{discord}
{\cal D}^{B\rightarrow A}(\rho_{AB})=\min_{\{\hat{\Pi}_B^j\}}[I(\rho_{AB})-J(\rho_{AB})].
\end{equation}

\subsection{Global discord}
The first multipartite measure of the QD we will consider is the global discord (GD)~\cite{Rulli}. The construction is a natural extension of a symmetrized version of the QD to $N$ particles. The symmetric form of the QD can be cast in terms of relative entropy $S(\rho_1||\rho_2)=\mathrm{Tr}[\rho_1\log_2\rho_1]-\mathrm{Tr}[\rho_1\log_2\rho_2]$ between two generic states $\rho_1$ and $\rho_2$ if we allow for bilateral measurements, $\hat\Pi^j_A\otimes\hat\Pi^k_B$~\cite{girolami}, to be performed. Thus
\begin{equation}
\begin{split}
\mathcal{D}^{AB}(\rho_{AB})=\min_{\{\hat \Pi^j_A\otimes\hat \Pi^k_B\}} & [S(\rho_{AB}||\hat\Pi(\rho_{AB}))\\
                                                  &       -\sum_{j=A,B}S(\rho_{j}||\hat{\Pi}_j(\rho_{j}))].
\end{split}
\label{Dsym}
\end{equation}
 where $\rho_j=\mathrm{Tr}_{i\neq	j}\left[\rho_T\right]$ is the reduced state of qubit $j$ and $\hat\Pi(\rho_{AB})=\sum_{j,k}(\hat\Pi^j_A\otimes\hat\Pi^k_B)\rho_{AB}(\hat\Pi^j_A\otimes\hat\Pi^k_B)$. This expression captures the quantum correlations associated with multi-local measurements. It should be noted that, except for a few special classes of states, this form of QD is not equivalent to another symmetrized version of discord where we take the maximum of Eq.~(\ref{discord}) attained by measuring $A$ and $B$ separately, i.e. $\mathcal{D}=\text{Max}[{\cal D}^{B\rightarrow A},{\cal D}^{A\rightarrow B}]$~\cite{girolami}.

Given a multipartite density matrix $\rho_{T}$, its GD is then defined as
\begin{equation}
\label{GQD}
\mathcal{G}_N(\rho_{T})=\min_{\{\hat\Pi^k\}}\left\{S\left(\rho_{T}||\hat\Pi(\rho_{T})\right)-\sum_{j=1}^N S\left(\rho_{j}||\hat\Pi_j(\rho_{j})\right)\right\},
\end{equation}
with  $\hat\Pi_j(\rho_{j})=\sum_{l}\hat\Pi_{j}^{l}\rho_{j}\hat\Pi_{j}^{l}$, $\hat\Pi(\rho_{T})=\sum_k \hat\Pi^k \rho_{T} \hat\Pi^k$, $\hat \Pi^k=\otimes^N_{l=1}\hat \Pi^{k_l}_{l}$, and $k$ stands for the string of indices $(k_1\dots k_N)$. As a global measure of the quantum correlations in a given multipartite state it was shown in~\cite{Rulli} the maximum value attainable is related to the dimensionality of the Hilbert space considered. The minimization  in Eq.~(\ref{GQD}) makes it an involved quantity to calculate. A more computationally efficient means to evaluate the GD is given in Ref.~\cite{campbellGD}. Furthermore, we find exploiting any symmetries present can greatly reduce the effort required in calculating Eq.~(\ref{GQD}).

\subsection{Genuine correlations in symmetric multipartite systems }
A definition of genuine quantum and classical correlations can be given starting from the generalization of the mutual information to $n$ parties \cite{genuine}: total correlations can be measured by  
\begin{equation}
T(\rho_T)=\sum_{j=1}^n S(\rho_j)-S(\rho_T).
\end{equation}
Genuine correlations are introduced in order to quantify all the correlations that cannot be accounted for considering any of the possible subsystems: 
a state of $n$ particles has genuine $n$-partite correlations if it is non-product in every bipartite cut~\cite{grudka}. According to this criterion, genuine total correlations $T^{(n)}(\rho_T)$ are defined 
as the distance, quantified through the relative entropy, between $\rho_T$ and the closest state with no $n$-partite correlations, that is, the closest state which is product along a bipartite cut  \cite{genuine}. For instance, total correlations of a tripartite density matrix whose parties are labeled as $i,j,k$ are given as
\begin{equation}
T^{(3)}(\rho)=\min_{k} [S(\rho_{ij})+S(\rho_{k})-S(\rho)].
\end{equation} 
 As a consequence of this definition, $T^{(n)}$ coincides with the  mutual information between two  complementary sub-parties. Then, we can apply definitions (\ref{CI}) and (\ref{discord}) associating these two sub-parties respectively to $A$ and $B$ and deriving in such a way a quantifier for classical ($J^{(n)}$) and quantum (${\cal D}^{(n)}$)  genuine correlations.  

In Ref. \cite{genuine}, it was shown that a consistent set of definitions for any level of separability can be done for pure states. The extension to the case of symmetric multipartite systems, which will be used in this paper, follows directly since there is no ambiguity in the choice of the subsystems.  

\section{Work extraction from $n$-identical batteries}
\label{sec:work}
Here we briefly recall the main ingredients of the protocol we consider, which is the same as the one studied in Refs.~\cite{Allahverdyan,alicki,hovna}. Our system is composed by a register of $n$ (identical) $d$-level quantum systems, each of them prepared in the initial state
\begin{equation}
\Omega=\sum_{k=0}^{d-1} p_k|k\rangle\langle k|.\label{eq1}
\end{equation}
The system is governed by the Hamiltonian 
\begin{equation}
h_0=H\otimes  1\!\!1\otimes\cdots \otimes  1\!\!1+\dots + 1\!\!1\otimes   1\!\!1\otimes\cdots, \otimes H\label{eq2}
\end{equation}
where the single-battery Hamiltonian is
\begin{equation}
H=\sum_{k=0}^{d-1} \varepsilon_k|k\rangle\langle k|.
\end{equation}
To the end of extracting work from $\Omega$, at $t=0$, an external potential $V(t)$ is switched on. The total Hamiltonian reads $h(t)=h_0+V(t)$. The process is cyclic if, at $t=\tau$, the external field is switched off and the system is returned to its initial configuration. The final state is then $\Omega(\tau)=U(\tau)\Omega U^\dag(\tau)$ where $U$ is the operator governing the time evolution.
The work extracted during the cycle is identified with the difference between the average energy before the field is switched on and the one after it is switched off:
\begin{equation}
{\cal W}={\rm Tr}\left[ \Omega h_0 \right]-{\rm Tr}\left[ \Omega(\tau) h_0\right].
\end{equation}
For infinite systems, the maximum value of ${\cal W}$ is obtained when the final state $\Omega(\tau)$ is the canonical Gibbs state
\begin{equation}
\Omega(\tau)=\Omega_{{\rm eq}}=\frac{e^{-\beta H}}{{\rm Tr}\left[e^{-\beta H}\right]},
\label{thermal}
\end{equation}
where  $\beta$ can be obtained observing that the von Neumann entropy cannot decrease during the cycle and that there exists a unique value of the inverse temperature such that $S(\Omega)=S(\Omega_{{\rm eq}})$. For finite systems, as the evolution is necessarily described by unitary operations and there is no dissipation, the thermal equilibrium is not reached. In this case,  ${\cal W}$ is maximized provided that $[h_0,\Omega(\tau)]=0$, $\Omega(\tau)$ and $\Omega$ share the same set of eigenvalues $\{p_i\}$, and finally, the eigenvalues of  $\Omega(\tau)$ are reversely ordered with respect to the ones of $h_0$  \cite{Allahverdyan}. 

A single-battery state $\Omega$ is called passive with respect to $H$ if no energy can be extracted from it during a cycle; it is called completely passive if $\Omega^{\otimes n}$ is passive with respect to $h_0$ for all $n$. Passive states are not necessarily completely passive, a remarkable exception being represented by qubits ($d=2$), due to the one-to-one correspondence between the entropy of the state and its ordered eigenvalues. 
This also implies that, working with qubits, the classical limit can never be exceeded, that is, the maximal wok extracted from $n$ batteries is equal to $n$ times the maximal single-battery work.
In Ref. \cite{alicki}, an example involving qutrits ($d=3$) is presented where also for $n=3$ there is no advantage in the multi-battery case. Actually, that result depends on the particular choice of the system and Hamiltonian parameters. Given $H$, it is in principle possible to engineer single-battery states that are optimal for any $d$ and for any finite $n$. However, the measure of the set of such  states  goes to zero in the thermodynamic limit.

A simple criterion to determine whether the classical limit is beaten or not is based on the use of classical correlations: the work extracted is  $n$ times the work that could have been extracted from a single battery if and only if the final state  is the tensor product of single-battery states. In fact, this condition is necessary because any product state could be obtained by local manipulation of the initial one; on the other hand, it is sufficient because local unitary operations map product states onto product states. Then, classical correlations can be used to measure the distance from the set of product states.

In Ref. \cite{hovna} it was shown that maximal work extraction can be reached without generating entanglement during the dynamics. In fact, the work extraction protocol described so far requires the implementation of swap operations  between Hamiltonian eigenstates. If the two eigenstates have $m$ different battery indices, the swap operation can be carried out  in $2m-1$ sequential steps, each of them only involving  a single battery index exchange. In this way, the protocol is accomplished and the state remains separable at any time. Alternatively, a single $m$-index swap can be implemented which turns out to be ``faster" than the previous one (the speed of the process being measured by the number of unitary operations used, however we will see that such an operation is also more optimal entropically). Faster processes can be accompanied by the presence of entanglement or not depending on the value of the parameters of the state and of the Hamiltonian, however we will show that quantum discord will always be produced. 

\section{Optimal work extraction}\label{owe}
Hovhannisyan \textit{et al.} demonstrated that the even if an entangling Hamiltonian is able to extract more work per copy than a local one, not necessarily does the density matrix become entangled during the extraction process \cite{hovna}. Indeed, multi-step strategies guarantee that the state remains separable at any time. However, the absence of entanglement does not imply the absence of quantum correlations. In the following we show, with the exception of the special case of $n$ identical two-level batteries (qubits), there exist no strategies such that the state can be described as a classical probability distribution, that is, the maximal extraction protocol cannot be implemented without dynamically producing discord. 

We start our discussion with two  cases that, in spite of their simplicity, contain all the ingredients we need for our purpose. 

\subsection{Two-qubit case}

Let us start considering the simplest situation, that is, the case of two qubits initially prepared in
\begin{equation}
\Omega=p_{0}^2|00\rangle\langle 00|+p_{0}p_{1}(|01\rangle\langle 01|+|10\rangle\langle 10|)+p_{1}^2|11\rangle\langle 11|,\label{omega}
\end{equation}
with $p_0+p_1=1$, in the presence of the Hamiltonian 
\begin{equation}
h_0=2 \epsilon_0 |00\rangle\langle 00|+2 \epsilon_1 |11\rangle\langle 11|+(\epsilon_0+\epsilon_1)(|01\rangle\langle 01|+|10\rangle\langle 10|),
\end{equation}
where $\epsilon_0<\epsilon_1$. As said in the previous section, the case of qubits is somewhat special, as single-battery passive states remain passive in the multi-battery scenario. This implies that the maximal work that can be extracted from a $n$-partite battery cannot exceed the one that could be obtained by processing $n$ separate batteries. Nevertheless, this case will be particularly instructive in elucidating the link between single-index and $m$-index swaps and the unavoidable generation of quantum correlations in the higher dimensional cases.

The state $\Omega$ introduced in Eq. (\ref{omega}) is active, that is, it is possible to extract work from it, provided that $p_{0}<p_{1}$. The amount of work extracted is maximum if, at the end of the cycle,
\begin{equation}
\Omega(\tau)=p_{1}^2|00\rangle\langle 00|+p_{0}p_{1}(|01\rangle\langle 01|+|10\rangle\langle 10|)+p_{0}^2|11\rangle\langle 11|.
\end{equation}
In this case, ${\cal W}_{\max}=2(\epsilon_1-\epsilon_0)(1-2 p_0)$. It is immediate to check that ${\cal W}_{\max}$ is twice the maximal work that could be extracted from a single battery. The reordering process can be performed either (i) in three steps or (ii) in one single step.  In the case (i) the procedure consists of first swapping, for instance, $|00\rangle$ and $|10\rangle$, then  $|10\rangle$ and $|11\rangle$, and, finally,  $|00\rangle$ and $|10\rangle$ again. In the case (ii) , there is a direct swap between $|00\rangle$ and $|11\rangle$. In this case it is immediate to see that the final entropic cost of these two protocols is the same, since the final states are identical. However, as protocol (i) requires three unitary operations each taking a time $\tau$ to be completed, it follows that the direct swap case is preferable as it can be implemented in a single operation without incurring any additional entropic cost. 

\begin{figure}[t]
\includegraphics[width=7cm]{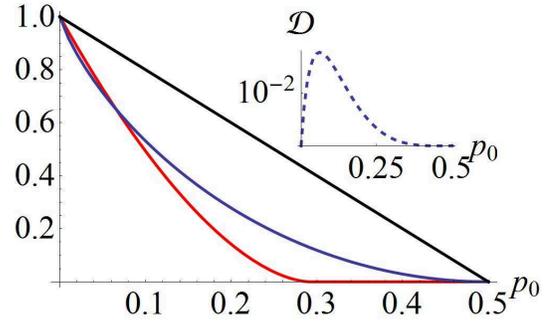}
\caption{  $ {\cal W}_{\max}$ (black), quantum discord (blue), and  entanglement of formation (red) as a function of $p_0$. The work extracted has been calculated assuming $\varepsilon_0=0$ and $\varepsilon_1=1$. Inset: quantum discord produced in the three-stage protocol. Its value is much smaller with respect to the single-stage case. The entanglement of formation is not plotted given that the density matrix is separable at all times.}
\label{fig1}
\end{figure}

As noticed  in Ref. \cite{hovna}, following the multi-step strategy, the state $\Omega(t)$ remains separable at all times, while, as a consequence of the direct swap, $\Omega(t)$ may or may not be entangled for some intermediate times between $t=0$ and $t=\tau$. That is, the presence of entanglement is somehow related to the speed of the extraction  process, however entanglement is not the only indicator of quantumness.  In Fig. \ref{fig1} we plot the maximum discord produced during the direct swap case, also showing with entanglement of formation and maximal work extraction. In the inset, we show the discord for the multi-step strategy.

From this example we learn that, even in the simplest scenario, implementing any swap operations has a cost in terms of quantum correlations between the sub-parts. Furthermore, the order of magnitude of the discord generated through direct swap is around two orders of magnitude larger than the discord produced following the three-stage strategy (inset). However, at this point we cannot draw any strong conclusions about work extraction, as a local transformation leading to the very same final state is available.

\subsection{Two qutrits}\label{twoqutrits}

The second case we analyze in detail is the one of two qutrits. The initial state is  
\begin{equation}
\Omega=(p_{0}|0\rangle\langle 0|+p_{1}|1\rangle\langle 1|+p_{2}|2\rangle\langle 2|)^{\otimes 2},
\end{equation}
with $p_0\le p_1\le p_2$, while the Hamiltonian is
\begin{eqnarray}
h_0&=&2 (\epsilon_0 |00\rangle\langle 00|+ \epsilon_1 |11\rangle\langle 11|+ \epsilon_2 |22\rangle\langle 22|)\nonumber\\
&+&(\epsilon_0+\epsilon_1)(|01\rangle\langle 01|+|10\rangle\langle 10|)\nonumber\\&+&(\epsilon_0+\epsilon_2)(|02\rangle\langle 02|+|20\rangle\langle 20|)\nonumber\\&+&(\epsilon_1+\epsilon_2)(|12\rangle\langle 12|+|21\rangle\langle 21|).
\end{eqnarray}
Let us assume, for the sake of clarity $\epsilon_0=0,\;\epsilon_1=0.579,\;\epsilon_2=1$ together with $p_0=0.224$. Let us also take $p_1$ (and $p_2$) as a free parameter with the constraints $p_0<p_1<p_2$ and $p_0+p_1+p_2=1$. This situation is closely related to the one considered in Ref. \cite{alicki}, the only difference being that we have swapped the value of $p_2$ with $p_0$ meaning our state is initially active. However we remark that making an active single-battery state passive just introduces a fixed amount of work that does not modify what can be extracted by employing nonlocal operations, as we are interested in the conditions under which the classical limit can be beaten this difference is essentially immaterial. Using a classical, i.e. local, protocol, the work extracted in the optimal case is, irrespective of $p_1$, 
\begin{equation}
{\cal W}_{{\rm cl}}=2(\epsilon_2-\epsilon_0)(p_2-p_0).
\end{equation} 
\begin{figure}[t]
\includegraphics[width=7 cm]{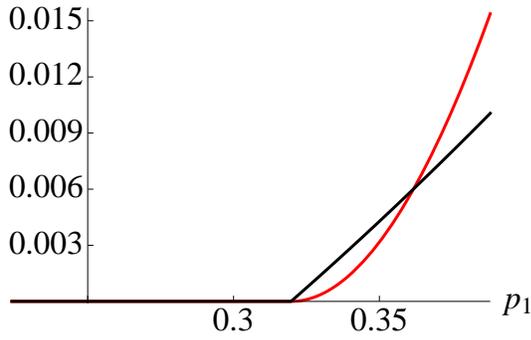}
\caption{  Classical correlations $J(\rho)$ (red) and difference between maximal work extracted and maximal work extracted using a classical protocol (black) for the two-qutrit case discussed in Sec. \ref{twoqutrits}.}
\label{fig2}
\end{figure}
Taking into account global unitary transformations, the final order of the eigenstates  depends on the value of $p_1$. There exists a threshold value $p_1^{\rm th}$ (obtained  imposing $p_1^2=p_0 p_2$) such that, for $p_1\le p_1^{\rm th}$, the maximum extractable work does not exceed the classical limit, as all we need is to swap $|0\rangle$ with $|2\rangle$ in any of the two qutrits. This is no longer true for $p_1 > p_1^{\rm th}$, where, besides the $|00\rangle\Leftrightarrow |22\rangle$ swap, we also need a further swap between $|11\rangle$ and either $|02\rangle$ or $|20\rangle$. The difference between maximal work extracted and maximal work extracted using a classical protocol is plotted in Fig.~\ref{fig2} together with the classical correlations left in the  state at the end of the cycle. The connection between these two quantities is clearly shown here. For $p_1<p_1^{\rm th}$, in agreement with \cite{alicki}, the maximum extractable work is equal to the classical limit. However, when $p_1>p_1^{\rm th}$ ($\approx0.322$ in the figure) the final state exhibits classical correlations, and we see that collective strategies allow one to extract more work from the two three-level batteries than in the classical limit. This then clearly indicates that the use of higher dimensional systems has advantages regarding how much work can be extracted.

While for two-qutrits we can in principle calculate the quantum discord directly, we notice that performing any of the swaps will result in the generation of only a single off-diagonal element in the density matrix. As pointed out in~\cite{hovna} we can map the two-qutrit state onto a two-qubit state which then allows us to use the results of the previous section to quantitatively and qualitatively examine the quantum discord. Full details are given in the appendix, however for clarity let us discuss the $|00\rangle\Leftrightarrow |22\rangle$ swap, the only off-diagonal element of the two-qutrit density matrix generated dynamically is $\ket{00}\bra{22}$ (and its Hermitian conjugate). We can then define a two qubit basis as $\{\ket{00},\ket{02},\ket{20},\ket{22}\}_{1,2}$. Normalizing the resulting matrix after projecting onto this basis gives us the mapped density matrix. A similar mapping can be done for any swap of indices. The relevance of the previous section should now be evident, the mapped form of the two-qutrit density matrices for any swap are analogous to those for the swapping operations performed in the two-qubit case. As shown in Fig.~\ref{fig1} these states always have quantum discord and we can take the quantitative value as a lower bound to the amount of discord present in the full two-qutrit state.

\subsection{General case: discord witness}
The previous two sections have given us all the necessary ingredients to discuss the generalization to $n$ $d$-level batteries. We will make use of the two measures of multipartite quantum discord introduced in Sec. \ref{sec:discord}. Our goal is to show that, beyond the quantification of discord, which can become a formidably complicated problem, it is possible to witness its presence by inspection of the shape of the density matrix. In other words, we are going to establish that  the creation of multipartite discord is necessary for maximal work extraction.

Let us consider the  general case of $n$ identical qudits initially  prepared in 
\begin{equation}
\Omega_{n,d}=(p_0|0\rangle\langle  0|+\dots +p_{d-1}|d-1\rangle\langle d-1|)^{\otimes n}.
\end{equation}
We only assume that $\Omega_{n,d}$ is an active state, that is, through a reordering process of its populations, it is possible to extract work from it. Among all the swap operations that, in general, need to be implemented, we focus on one of them.  Let $|\alpha\rangle$ and $|\beta\rangle$ be the states to be swapped.
 During the swap, the density matrix is 
\begin{equation}
\begin{split}
\Omega_{n,d}(t)=&\textrm{diag}(\Omega_{n,d}^{(i)})+c_{\alpha,\alpha}(t)|\alpha\rangle\langle \alpha|+c_{\beta,\beta}(t)|\beta\rangle\langle \beta| \\
                            &+(c_{\alpha,\beta}(t)|\alpha\rangle\langle \beta|+h.c.),\label{omegand}
\end{split}
\end{equation}
where $\Omega_{n,d}^{(i)}$ are all the eigenvalues of $\Omega_{n,d}$ whose eigenvectors are not involved in the swap, and where the shape of the time-dependent coefficients depends on $V(t)$. 

According to the definition given in Ref. \cite{genuine}, genuine total correlations ${\cal D}^{(n)}$ can be calculated considering any possible bipartite cut along the system and taking the minimum mutual information among them. Then, genuine quantum discord is the quantum part of a bipartite mutual information and can be calculated following the general rules used for bipartite systems. 
Let us first assume that all the indices of $|\alpha\rangle$ and $|\beta\rangle$ are different, that is, let us work in the direct swap case. Considering \textit{any} bipartition $\{a,b\}$, the density matrix can always be written as 
\begin{equation}
\begin{split} 
\Omega&_{n,d}(t)=\sum_{i \in a,\;j \in b } \nolimits'  c_{ij}|i,j\rangle\langle i,j| +c_{\alpha,\alpha}(t)|\alpha_a,\alpha_b\rangle\langle \alpha_a,\alpha_b|  \\
                         &+c_{\beta,\beta}(t)|\beta_a,\beta_b\rangle\langle \beta_a,\beta_b|+(c_{\alpha,\beta}(t)|\alpha_a,\alpha_b\rangle\langle \beta_a,\beta_b|+h.c.),
\end{split}
\label{bipartition}
\end{equation}
 where $i$ and $j$ run over the whole subspace of the respective partition and where the prime 
 indicates that the elements $|\alpha\rangle\equiv |\alpha_a,\alpha_b\rangle$ and  $|\beta\rangle\equiv |\beta_a,\beta_b\rangle$ are excluded from the sum.  Given that the state of Eq. (\ref{bipartition}) has never a quantum-classical form (there is always a non-diagonal part in any of the two sub-parties), we conclude that genuine discord is present with certainty at any time $0<t<\tau$.

Let us now consider that case of partial swap, that is, the case where $|\alpha\rangle$ and $|\beta\rangle$ share only a sub-set of $m < n$   battery indices.  Under this hypothesis, there is a natural bipartition between the set of common indices $\{x\}$ and the complementary set $\{\bar x\}$. Assuming $\{a\equiv x\}$ and $\{b\equiv \bar x\}$, which implies $\alpha_a=\beta_a$, it is immediate to see that the state has a quantum-classical form:
\begin{equation}
\begin{split}
\Omega&_{n,d}(t)= \sum_{i \in a,\;j \in b } \nolimits'  c_{ij}|i,j\rangle\langle i,j|+|\alpha_a\rangle\langle \alpha_a|\otimes \\
&\left[ c_{\alpha,\alpha}(t)|\alpha_b\rangle\langle \alpha_b| +c_{\beta,\beta}(t)|\beta_b\rangle\langle \beta_b|+(c_{\alpha,\beta}(t)|\alpha_b\rangle\langle \beta_b|+h.c.)\right].
\end{split}
\label{bipartition2}
\end{equation} 
This implies that, in accordance with Ref. \cite{genuine}, genuine quantum correlations are zero. Nevertheless, total quantum correlations are not vanishing. Indeed, considering the reduced density matrix taken tracing out the  $m$ vectors that are left unchanged, we obtain a state that is formally identical  to the state of Eq. (\ref{bipartition}) (in other words, the structure is preserved). This is enough to say that there are $m$-partite quantum correlations.

As proven in Ref. \cite{witness}, the presence of global discord in a density matrix $\rho$ can be witnessed by calculating the commutator between $\rho$ itself and the tensor product of the marginals $\tilde{\rho}=\rho_1\otimes\rho_2\otimes\cdots\otimes \rho_n$ ($\rho_i=Tr_{j\neq i}[\rho]$). In fact, if ${\cal C}=[\rho,\tilde{\rho}]\neq 0$, the state is not classical, that is, it is not possible to find a local measurement basis such that $\rho$ is left unchanged. Let us apply this criterion to our case. Under total swap,  $\tilde{\rho}$ is diagonal. As it can be explicitly verified, the commutator ${\cal C}$ has nonvanishing matrix elements:
\begin{equation}
\langle \alpha|{\cal C}|\beta\rangle=-\langle \beta|{\cal C}|\alpha\rangle=\langle \beta|\rho|\alpha\rangle\left(\langle \beta|\rho_1|\beta\rangle-\langle \alpha|\rho_1|\alpha\rangle\right).
\end{equation}
As a consequence, there is always global discord during the swap operation, unless $\langle \beta|\rho_1|\beta\rangle=\langle \alpha|\rho_1|\alpha\rangle$. 
Actually, this condition  can only occur for pairs of qubits  ($n=d=2$) and for $\omega t=\pi/4$, but in this simple case the discord can be explicitly evaluated without recurring to a witness.

\begin{figure}[h!]
{\bf (a)} \\
\includegraphics[width=6cm]{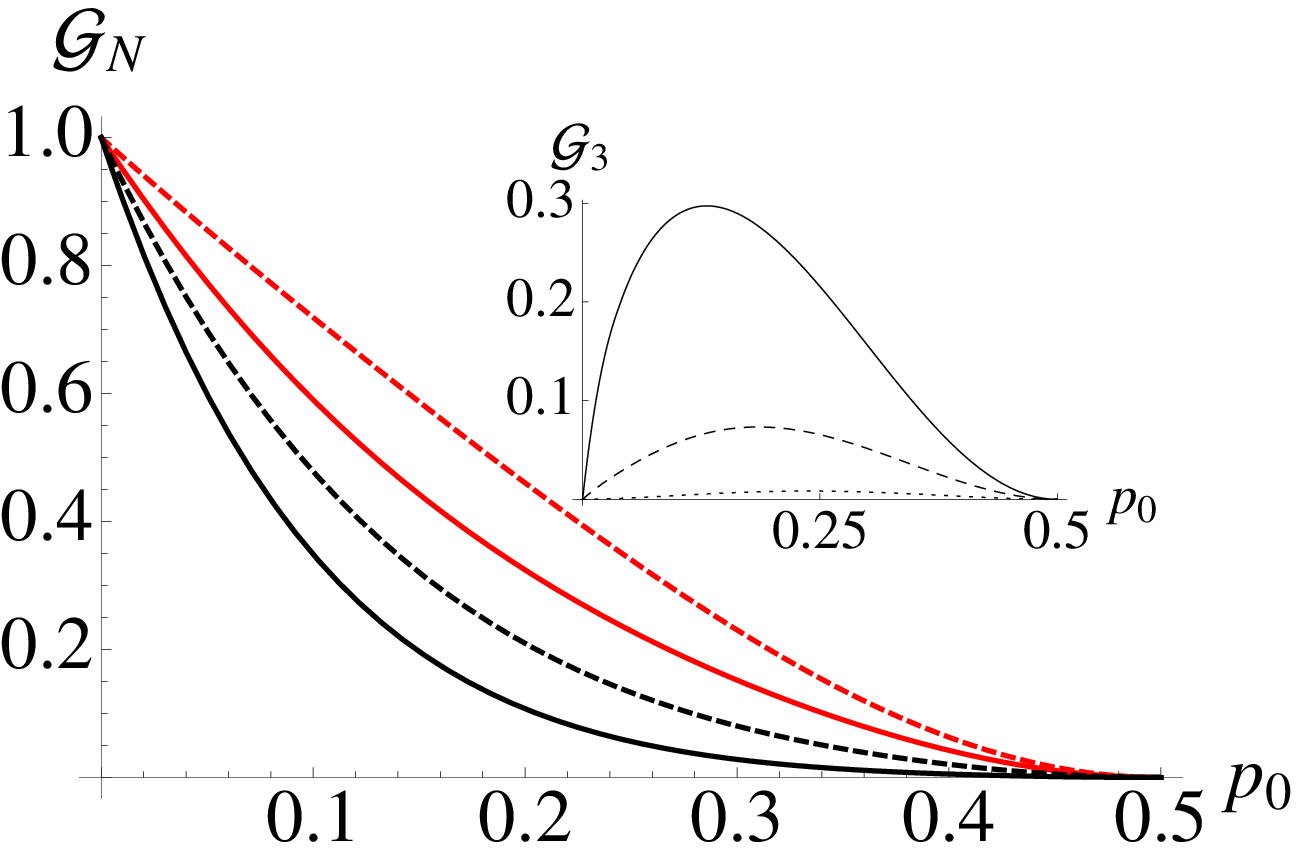}\\
{\bf (b)} \\
\includegraphics[width=6cm]{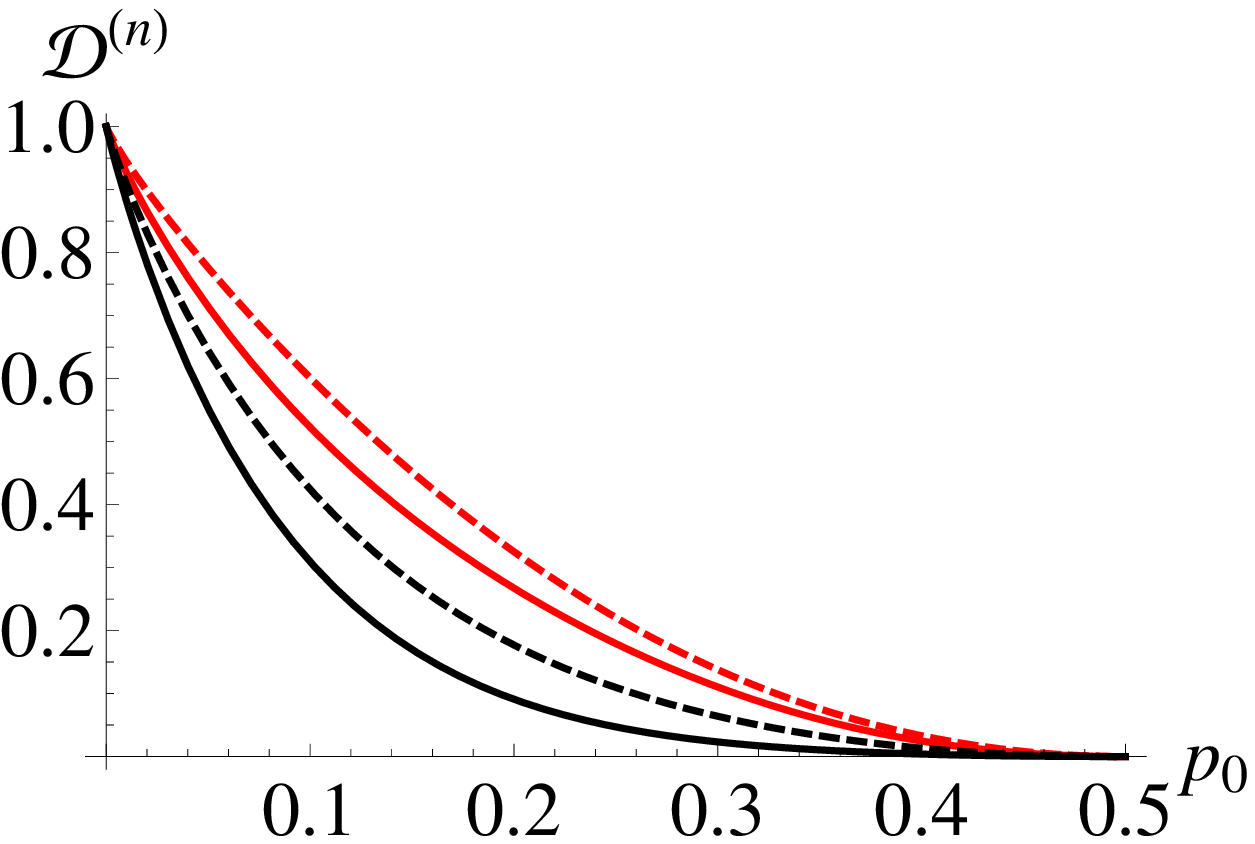}\\
{\bf (c)} \\
\includegraphics[width=6cm]{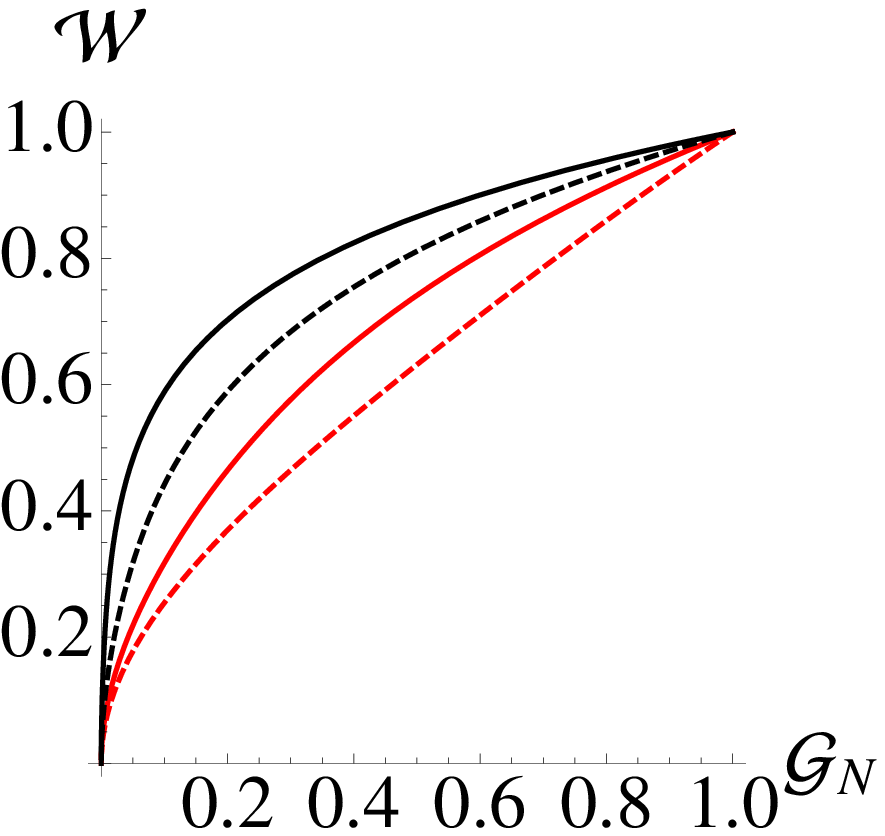}
\caption{  Maximum amount of GD {\bf (a)} and GC {\bf (b)} produced in a $n$-qubit system as a function of $p_0$ for $n=3$ (red dashed),  $n=5$ (red solid), $n=7$ (black dashed),  $n=10$ (black solid). The initial state is the one given in Eq. (\ref{omegand}) and the work is extracted swapping the populations $p_0^n$ and $p_1^n$ of  $|0,0,\dots,0\rangle$ and $|1,1,\dots,1\rangle$. Inset of {\bf (a)}: Global Discord for $n=3$ where the swap is performed in 5 stages (i) $\ket{000}\Leftrightarrow \ket{100}$, (ii) $\ket{100}\Leftrightarrow \ket{110}$, (iii) $\ket{110}\Leftrightarrow \ket{111}$, (iv) $\ket{100}\Leftrightarrow \ket{110}$, and (v) $\ket{000}\Leftrightarrow \ket{100}$. While no GC (and hence no entanglement) is created, we see the maximum GD generated during (i) and (v) dotted (lowest) curve, (ii) and (iv) dashed (second) curve, and (iii) solid (top) curve of the inset. {\bf (c)} Maximum dynamically generated GD versus extractable work from a direct swap of the largest and smallest eigenvalues for $n=3$ (red dashed),  $n=5$ (red solid), $n=7$ (black dashed),  $n=10$ (black solid) in the qubit case and we have taken $\varepsilon_0=0$ and $\varepsilon_1=1$.}
\label{fig3}
\end{figure}

The proof that ${\cal C}\neq 0$ also in the case of partial swap can be done in two steps. Let us consider a $n$-partite system, a generic bipartite cut $\{a\equiv 1,2,\cdots,k;b\equiv k+1,k+2,\cdots n\}$, and assume $[\rho,\rho_a\otimes\rho_b]\neq 0$, that is, let us assume that there is global discord (according to Eq. (\ref{Dsym}) between the two sub-parties.
Then, by definition, there must be at least the same amount of  global discord when considering all the $n$ parties separately. This can be deduced observing that  in Eq. (\ref{GQD}) the measurement is performed using local operators. If we enlarge the set of possible projectors to multipartite entangled states,  extended to  either  $\{a\}$ or  $\{b\}$, we get at worst the same  minimum that would be  obtained considering all the $n$ parties separately:
\begin{equation}
\mathcal{G}_N(\rho_{a\equiv 1,2,\cdots,k,b\equiv k+1,k+2,\cdots n})\le \mathcal{G}_N(\rho_{1,2,\dots,n})
\end{equation}
 As a second and final step, we need to prove that there is bipartite global discord in $\Omega_{n,d}(t)$.  As a consequence of the condition ${\cal C}=0$, the reduced states of the system must obey $\rho_j=\sum_n p_n \Pi^n_j$, that is, they are required to be diagonal in a measurement basis \cite{witness}. This condition is clearly violated by $\Omega_{n,d}(t)$ in Eq. (\ref{bipartition2}), as the trace over party $a$ has always nondiagonal elements.

\subsection{General case: quantification of discord}
Finally we notice that in the $n$-qubit case for a direct swap of the largest and smallest eigenvalues we can compute the maximum global discord dynamically generated analytically,
\begin{equation}
\begin{split}
\mathcal{G}_N^\text{max}=&p_0^N \log_2 p_0^N+p_1^N \log_2 p_1^N  \\
                                             &-\left(p_0^N+p_1^N\right) \log_2 \left(\frac{1}{2} \left[p_0^N+p_1^N\right]\right).
\end{split}
\label{analyticGQD}
\end{equation}
In virtue of the mapping described in the Appendix this then becomes a lower bound for arbitrary dimensional systems. In Fig.~\ref{fig3} {\bf (a)} and {\bf (b)} we plot the GD and GC for $n=3,5,7$ and 10 qubits, with a similar behavior holding for larger $n$. For all values of $p_0$ the state has non-zero discord, and as such we can extrapolate that the same behavior holds for $d$-dimensional systems. For the partial swap case determining a closed expression for the GD or applying the witness criteria is significantly more involved due to the form of the state even for qubits. However, for small systems we can still directly calculate the GD. The inset shows the case for $n=3$ qubits with quantum correlations always present. Finally, in Fig.~\ref{fig3} {\bf (c)} we show the one-to-one behavior between the extractable work under a direct swap of the largest and smallest eigenvalues and the global discord for $n=3,5,7$ and 10 qubits. As a consequence of the qudit-qubit mapping discussed the Appendix, the qualitative behavior found in  Fig.~\ref{fig3} {\bf (c)} would be found also in higher dimensions.

\section{Conclusions}
\label{sec:conclusions}
We have studied the problem of work extraction in a system consisting of $n$ independent $d$-dimensional batteries under the action of global operations. Starting with the relevant pedagogical cases of two qubits and two qutrits we have shown that, in general, these operations cannot be implemented without generating quantum discord. Furthermore, when considering $d$-dimensional batteries we have noticed that the maximal amount of extractable work is related to the correlations present in the final state. If the state is fully factorizable then we found the amount of extractable work is the same as the best possible classical scheme, i.e. local operations independently applied to each battery.
On the other hand, having classical correlations in the final state means that more work has been extracted with respect to any classical protocol.
Regardless, we find in general the swapping operations leading to extractable work require discord to be generated dynamically and there appears to be a one-to-one relation between the presence of discord and the ability to extract work. Finally, most of the actual extractable work is related to the swapping of the largest and smallest energy levels of the initial system. In this instance, it can be clearly seen that using a single swapping operation is preferable, as it does not have any associated additional entropic cost compared to the protocol where the state remains separable at all times due to performing the swap in $2n-1$ steps. In this case, while for some parameter values the state may be separable, there is always non-zero quantum discord. 

It is worth mentioning here that  our work is focused on a particular scheme based on swap gates. In principle,  generic unitary operations in generic protocols should be considered in order to assess the necessity of dynamical  production of quantum correlations.
We expect our results to lead to further study into the role that correlations, both classical and quantum, play in important thermodynamic processes.

\acknowledgments
The authors are indebted to Gabriele De Chiara, Marcus Huber, Mauro Paternostro, and Vlatko Vedral for invaluable discussions. SC is funded through the EU Collaborative Project TherMiQ (Grant Agreement 618074), GLG acknowledges financial support from Compagnia di San Paolo.

\renewcommand{\theequation}{A-\arabic{equation}}
\setcounter{equation}{0}  
\section*{APPENDIX - Mapping qudits to qubits}  
In this appendix we outline the procedure to map our qudits to qubits when only a single off-diagonal element is present in the density matrix. Let us consider the two-qutrit state that results in the direct swap of the largest and smallest eigenvalues considered in the main body of the text
\begin{widetext}
\begin{equation}
\varrho = 
\left(
\begin{array}{ccccccccc}
p_0^2\cos^2(t)+p_2^2 \sin^2(t) & 0 & 0 & 0 & 0 & 0 & 0 & 0 & i(p_0^2-p_2^2)\cos(t)\sin(t) \\
0 & p_0 p_1 & 0 & 0 & 0 & 0 & 0 & 0 & 0 \\
0 & 0 & p_0 p_2 & 0 & 0 & 0 & 0 & 0 & 0 \\
0 & 0 & 0 & p_0 p_1 & 0 & 0 & 0 & 0 & 0 \\
0 & 0 & 0 & 0 & p_1^2 & 0 & 0 & 0 & 0 \\
0 & 0 & 0 & 0 & 0 & p_1 p_2 & 0 & 0 & 0 \\
0 & 0 & 0 & 0 & 0 & 0 & p_0 p_2 & 0 & 0 \\
0 & 0 & 0 & 0 & 0 & 0 & 0 & p_1 p_2 & 0 \\
-i(p_0^2-p_2^2)\cos(t)\sin(t) & 0 & 0 & 0 & 0 & 0 & 0 & 0 & p_0^2\cos^2(t)+p_2^2 \sin^2(t)
\end{array}
\right)
\end{equation}
\end{widetext}
As the state has only a single off-diagonal element at $\ket{00}\bra{22}$ we can express the state in a reduced Hilbert space spanned by 2 qubits using the basis $\{ \ket{0}, \ket{2}\}_1$ and $\{ \ket{0}, \ket{2}\}_2$. We simply project our state onto this basis to obtain the sub normalized two-qubit state, $\rho_2$, with elements
\begin{eqnarray*}
&\bra{00} \varrho \ket{00} = p_0^2\cos^2(t)+p_2^2 \sin^2(t), \\
&\bra{02} \varrho \ket{02} = p_0 p_2, ~~~
\bra{20} \varrho \ket{20} = p_0 p_2, \\
&\bra{22} \varrho \ket{22} = p_2^2\cos^2(t)+p_0^2 \sin^2(t), \\
&\bra{00} \varrho \ket{22} =(\bra{22} \varrho \ket{00})^* = i(p_0^2-p_2^2)\cos(t)\sin(t),
\end{eqnarray*}
with all other elements zero. Which is then normalized simply dividing by the trace, $\text{Tr}[\rho_2]=p_0^2+p_2^2$,
\begin{equation}
\rho=\frac{1}{p_0^2+p_2^2} \rho_2.
\end{equation}
While it may appear we have simply thrown away any terms involving $p_1$ we should remember that the values of $p_0$ and $p_2$ are constrained under the normalization condition of the original state (i.e. $p_0+p_1+p_2=1$). For $n$ qutrits, with the same single off diagonal at $\ket{0^{\otimes n}}\bra{2^{\otimes n}}$, we will have the same coherence term appearing in the mapped $n$-qubit state. By performing the projections onto the $n$-qubit basis we see the only terms that are kept in the populations are those involving $p_0^n p_2^{n-k}$. It is easy to check that the order they appear in the mapped density matrix is precisely the same as appears in the $n$-qubit cases.

Such an approach can be performed for an arbitrary swap of any system. Assuming the single off diagonal element appearing in the density matrix is $\ket{\alpha_1 \dots \alpha_n}\bra{\beta_1 \dots \beta_n}$, we choose our new $n$-qubit bases to be $\{\ket{\alpha_i}, \ket{\beta_i}\}_i$. The mapped density matrix is obtained by normalizing the resulting matrix from projecting onto these bases.

\end{document}